\documentclass[aps,pra]{revtex4}

\usepackage{german,epsfig,pstricks}

\usepackage{amsmath,amssymb}

\DeclareTextSymbolDefault{\i}{OT1} % dotless i

\def\NN{{\rm I\kern-.25em N}}
\def\ZZ{{\sf Z\kern-.44em Z}}
\def\CC{{\rm\kern.24em \vrule width.04em height1.46ex depth-.07ex
\kern-.30em C}}
\def\trace{{\rm tr}\;}
\def\RR{{\rm
         \vrule width.04em height1.58ex depth-.0ex
         \kern-.04em R}}
\def\id{{\rm 1\kern-.22em l}}
\def\DJo{$\;$\kern-.4em \hbox{D\kern-.8em\raise.15ex\hbox{--}\kern.35em okovi\'c}}

\newcommand{\braket}[2]{\langle #1 | #2 \rangle}
\newcommand{\beq}{\begin{equation}}
\newcommand{\beqa}{\begin{eqnarray}}
\newcommand{\eeq}{\end{equation}}
\newcommand{\eeqa}{\end{eqnarray}}
\newcommand{\bra}[1]{\left\langle #1 \right |}
\newcommand{\ket}[1]{\left | #1 \right\rangle}
\newcommand{\nbeqa}{\begin{eqnarray*}}
\newcommand{\neeqa}{\end{eqnarray*}}
\newcommand{\bigfrac}[2]{\mbox {${\displaystyle \frac{ #1 }{ #2 }}$}}
\newcommand{\Matrix}[2]{\left( \begin{array}{#1} #2 \end{array}
  \right)}

\newtheorem{pdef}{Definition}[section]
\newtheorem{Corollary}{Corollary}[section]
\newtheorem{Theorem}{Theorem}[section]
\newenvironment{defi}[1]{\begin{pdef} {\em  (#1)} \begin{quote}}{\end{quote}\end{pdef}}
\newenvironment{since}{${}^{}$\\ {\bf Proof}:  \begin{quote}\begin{em}}{\begin{flushright}
                   {\bf q.e.d.}\end{flushright}\end{em}\end{quote} }

\begin{document}

\title{The invariant-comb approach and its relation to the balancedness of multipartite entangled states}
\author{Andreas Osterloh}
\affiliation{Fakult{\"a}t f{\"u}r Physik, Campus Duisburg, 
Universit{\"at} Duisburg-Essen, Lotharstr. 1, 47048 Duisburg, Germany.}
\author{Jens Siewert}
\affiliation{Departamento de Qu\'{\i}mica F\'{\i}sica, Universidad del Pa\'{\i}s Vasco -
             Euskal Herriko Unibertsitatea, Apdo. 644, 48080 Bilbao, Spain}
\affiliation{Ikerbasque, Basque Foundation for Science, Alameda Urquijo 36, 48011 Bilbao, Spain} 
\affiliation{Institut f\"ur Theoretische Physik, Universit\"at Regensburg, D-93040 Regensburg, Germany} 

\begin{abstract}

The invariant-comb approach is a method to construct entanglement measures 
for multipartite systems of qubits.
The essential step is the construction
of an antilinear operator that we call {\em comb} in reference to the
{\em hairy-ball theorem}. 
An appealing feature of this approach is that for
qubits (or spins 1/2) the combs are automatically invariant under $SL(2,\CC)$,
which implies that the  obtained invariants are entanglement monotones by 
construction.
By asking which property of a state determines whether or not it is detected
by a polynomial $SL(2,\CC)$ invariant we find that it is  the presence of a {\em balanced part} that
persists under local unitary transformations.
We present a detailed analysis for the maximally entangled states detected by 
such polynomial invariants, which leads to the  concept of {\em irreducibly balanced} states. 
The latter indicates a tight connection with SLOCC classifications  of qubit entanglement.  \\
Combs may also help to define measures
for multipartite entanglement of higher-dimensional subsystems. However, for higher spins
there are many independent combs such that it is non-trivial to find an invariant one.
By restricting the allowed local operations to rotations of the coordinate system
(i.e. again to the $SL(2,\CC)$) we manage to define a unique extension of the 
concurrence to general half-integer spin with an analytic convex-roof expression for mixed states.
\end{abstract}

\maketitle

\section{Introduction}

Entanglement is one the most counterintuitive features of 
quantum mechanics~\cite{Bell87} 
of which there is only rather incomplete knowledge until 
now.
We will define a quantum mechanical state of distinguishable particles
as having no global entanglement with respect to a given 
partition ${\cal P}$ of the system if and only if ({\em iff}) it can be written as a 
tensor product of the parts of some subpartition 
of it; a state of indistinguishable particles 
we call not globally entangled {\em iff} it can be written as the proper
symmetrization due to the particles' statistics of such a
tensor product\cite{Ghirardi02}. 
 The many different ways of partitioning a physical system already
imply that 
there are many families of entanglement in multipartite systems
or even bipartite systems with many inner degrees of freedom.
The concept of entanglement instead remains unaltered. 
Having agreed upon how to decompose the physical system
such that every quantum state can be expressed as 
a superposition of tensor products of states of its parts,
the entanglement of its components follows the definition given above.

In order to be more specific, 
let ${\cal H}_i$ be the $i$th {\em local} Hilbert space of 
some partition of the total Hilbert space 
${\cal H}=\prod^\otimes_{i\in I} {\cal H}_i$. In this case the partition would
be ${\cal P}:=\{\,{\cal H}_i\; ;\; i\in I\,\}$;
If $I_1\cup I_2=I$, then ${\cal P}_{sub}:=
	\{\,{\prod}^\otimes_{i\in I_1}{\cal H}_i 
	   ,{\prod}^\otimes_{i\in I_2}{\cal H}_i \,\}$
is a two-elemental subpartition of ${\cal P}$.
We call an operator on ${\cal H}$
{\em ${\cal P}$-local} iff it is a tensor product with respect to 
the partition ${\cal P}$. When it is clear from the context what 
the partition is, we will just use the term {\em local}.

While for two qubits there is only one type of entanglement,
it has been noticed rather early that starting from the three-qubit
case there is more than one class of entanglement~\cite{Duer00}.
That is, for more than two parties there are different classes
of states which are not interconvertible using only
Stochastic Local Operations and Classical 
Communication (SLOCC)~\cite{MONOTONES,SLOCC,Duer00}.
Due to this complication, a key question  which has
not been answered yet in the frame of a general theory is (despite
considerable efforts, see e.g., Refs~\cite{Duer00,Jaeger,Albeverio01,Klyachko2002,VerstraeteDMV02,VerstraeteDM02,Miyake02,Luque02,VerstraeteDM03,Briand03,Miyake03,BriandLT03,Miyake04,Lamata07,DoOs08,Bastin09})
how to classify, to detect, and to quantify multipartite pure-state 
entanglement in a sensible and physically justified way.
Nice overviews of the state of the art are given 
in Refs.~\cite{Horodecki07,Plenio07}.
Further, an initial analysis of systems of two and three qutrits can be found
in Refs.~\cite{Cereceda03,Briand03}, and the entanglement sharing
properties for qudits have been studied in Ref.\cite{Dennison01}.
An interesting account on activities with respect to higher local dimension
is given in Ref~\cite{Wootters01}.
Several collective multipartite measures for entanglement have been proposed
for pure 
states~\cite{Wallach,Wong00,Barnum03,HEYDARI04,Scott04,Mintert05,Love06}. 
Since these approaches 
have no control about how various classes of entanglement will be 
weighted in such a measure, the decision in favor of one specific
collective entanglement measure is arbitrary unless we 
gain significantly better understanding of the struture of entanglement itself.

In order to get additional insight,
class-specific entanglement measures provide one research line to pursue.
As an example of such a measure for three qubits, 
the three-tangle has been derived~\cite{Coffman00}
as the unique measure that discriminates the two distinct classes
of entanglement in three-qubit systems. It separates $W$ states
from the genuine class of three-qubit entanglement represented by 
the Greenberger-Horne-Zeilinger (GHZ) state. 
As well as its two-qubit counterpart, 
the concurrence~\cite{Hill97,Wootters98,Abouraddy01},
the three-tangle is a  polynomial $SL(2,\CC)$ invariant.
A procedure for the construction of similar class-specific
entanglement quantifiers has been developed in Refs.~\cite{OS04,OS05}.
It has been systematically analyzed for four and five qubit systems
in Ref.~\cite{DoOs08}. 

These measures combine a variety of desirable
properties of class-specific quantifiers of global entanglement.
Given a specific partition ${\cal P}$, a 
{\em measure for genuine $n$-tanglement} 
${\cal E}_g:{\cal H}\longrightarrow [0,1]$ 
should satisfy the following requirements
\begin{quotation}
\item[(i)]{${\cal E}_g[\Pi]=0$ for all pure product states $\Pi$, 
relative to the partition ${\cal P}$}
\item[(ii)]{
invariance under ${\cal P}$-local unitary transformations}
\item[(iii)]{the entanglement monotone property~\cite{MONOTONES}; i.e.
the measure must not increase (on average)
under {\em Stochastic Local Operations and Classical Communication} 
(SLOCC)\cite{SLOCC}}
\item[(iv)]{invariance under permutations of the 
${\cal P}$-local Hilbert spaces\cite{Coffman00} is desirable.}
\end{quotation}
Clearly, the most basic requirement appears to be the condition (i).
This is accomplished by what has been 
termed  a {\em filter} in Refs.~\cite{OS04,OS05,DoOs08}, that is, 
an operator that ``filters out'' all product states, in the sense that it has 
zero expectation value for them. In other words, the filter image
of any product state must be orthogonal to the original product state.
The filters are built from so-called comb operators for one- and 
two-copy single qubit states (see below).
This approach appears appealing since, interestingly, for qubits 
it automatically implies $SL$ invariance and thus
the monotone property~\cite{VerstraeteDM03}. 
Consequently, also (ii) and (iii) are satisfied.
Nevertheless, already for five qubits there is a large number of
such measures 
(even after imposing condition (iv)~\cite{DoOs08}) such that
one would like to understand their essence more deeply, and to
reduce their number on physical grounds.

In this paper we analyze what is common to the states that are
detected by the $SL(2,\CC)$ invariant operators. We find in general that only the
{\em balanced} part of a state is measured, a property that had already
been noticed for the three-qubit case in Ref.~\cite{Coffman00},
in terms of a geometrical interpretation.
On the other hand it is known that the modulus of
a polynomial invariant assumes its maximum on the set of stochastic
states~\cite{VerstraeteDM03}. The combination of these characteristics
leads us to the concept of irreducibly balanced states. 
(We would like to mention that Klyachko {\em et al.} have discussed
maximum multipartite entanglement and its relation to quantification
of multipartite entanglement from a different, very interesting
perspective, see Refs.~\cite{Klyachko2002,Klyachko2004} and references
therein.)
We study
various interesting properties of irreducibly balanced states as we are convinced 
that their investigation might give further insight into the nature 
of multipartite entanglement.

As the basis of all this discussion is the invariant-comb approach,
it is interesting to ask whether there is any possibility to extend
the method to systems with local Hilbert space dimension larger
than $2$.  We discuss some basic aspects of such an extension.

The structure of the paper is as follows.
The invariant-comb approach for qubits~\cite{OS04,OS05} is summarized
in section \ref{approach};  
section~\ref{combs} introduces the main concepts, notations,
and the elements that eventually build up the (filter) invariants. 
In section \ref{unknownfilters} we exemplarily write down filters
for up to six qubits and discuss some elementary properties of them.
Section \ref{maxent} is devoted to the 
notion of states with maximal $n$-qubit entanglement.
Interestingly, a central prerequisite of such maximally entangled states
can be connected with the concept of {\em balancedness} in section~\ref{balancedness}. 
In section \ref{eval} we show that it is precisely the
balanced part that is detected by the $SL(2,\CC)$ invariant (filter) operators. 
In section \ref{irredbal}, this leads us to the 
definition of irreducibly balanced states and the investigation of
their properties.
Finally, we discuss the possibilities for an application 
of the invariant-comb approach to non-qubit systems and general partitions
in section \ref{gen}. First we focus on the entanglement of blocks
of qubits (part \ref{compounds}), while in part \ref{spinS} a measure
for bipartite entanglement, subject to a certain restricted 
class of local operations,  is derived for general half-integer spin
pure and mixed states.
In the last section, we present our conclusions.

\section{Polynomial SL(2,$\CC$) invariants for multipartite entangled states 
          -- the invariant comb approach}
\label{approach}

\subsection{The concept of combs and filters}\label{combs}

The fundamental concept of the method and the basis of the construction of 
polynomial invariants is the {\em comb}.
A comb $A$ is a local antilinear operator with 
zero expectation value for all states of the local Hilbert spaces, 
that is
\beq\label{def:Toeter}
\bra{\psi} A_i \ket{\psi}=\bra{\psi} L_i {\mathcal C}\ket{\psi}
	=\bra{\psi} L_i \ket{\psi^*}\equiv 0\quad 
\mbox{on } {\cal H}_i\; ,
\eeq
where ${\mathcal C}$ is the complex conjugation in the computational basis
\nbeqa
\ket{\psi^*}&:=&{\mathcal C}\ket{\psi}\equiv 
{\mathcal C}\sum_{j_1,\dots,j_q=1}^1 \psi_{j_1,\dots,j_q}\ket{j_1,\dots,j_q}\\
&=&\sum_{j_1,\dots,j_q=1}^1 \psi_{j_1,\dots,j_q}^*\ket{j_1,\dots,j_q}\; .
\neeqa 
We call $L_i$ the linear operator associated to the comb $A_i$.
For simplicity we abbreviate
\beq
\bra{\psi} L_i {\mathcal C}\ket{\psi}=:(\!( L_i )\!)
\eeq
(throughout this article, there will be no ambiguities whether we mean
linear or antilinear expectation values).
The requirement to have vanishing expectation values for an arbitrary single qubit state
clearly cannot be accomplished by any linear operator (it would be identically zero);
but it is amenable to antilinear operators.
The idea is to identify a sufficiently large set of combs in order
to construct the desired filter operators that satisfy 
all the requirements (i) - (iv) listed above.
It is worth noticing that a filter constructed exclusively from
combs automatically is invariant under ${\cal P}$-local unitary 
transformations if the combs are. Even more, it is invariant with respect to 
the complex extension of the corresponding unitary group, 
which is isomorphic to the corresponding special linear group. Since the latter mathematically represents the 
non-projective LOCC
operations~\cite{Duer00,Jaeger},
SL-invariants are entanglement monotones by construction.
This follows from an important theorem by Verstraete {\em et al.}:
{\em Consider a linearly homogeneous positive function of a pure
(unnormalized) state $M(|\psi\rangle\!\langle\psi |)$ that 
remains invariant under determinant 1 SLOCC operations} 
($SL$ operations - authors' remark). 
{\em Then $M(|\psi\rangle\!\langle\psi |)$ is an entanglement monotone.}
(For the proof, cf.~Ref.~\cite{VerstraeteDM03}.) It is evident that
any polynomial invariant can be turned into a linearly homogeneous
function of $|\psi\rangle\!\langle\psi |$ by applying the appropriate
inverse power. In order to avoid misunderstandings we emphasize that 
not every function of SL-invariants (which still is an SL invariant)
can be an entanglement monotone; it is not even clear whether homogeneity
of arbitrary degree implies the monotone property.

The main part of this work focuses on multipartite registers of
$n$ qubits,  i.e., $n$ spins 1/2. Then, the local Hilbert space is 
${\cal H}_i=\CC^2=:\mathfrak{h}$ for all $i\in\{1,\ldots,n\}$. We will need the Pauli matrices
\beq\label{def:Paulis}
\sigma_0:=\id_2=\Matrix{cc}{1&0\\0&1}\, ,\; 
\sigma_1:=\sigma_x=\Matrix{cc}{0&1\\1&0}\, ,\; 
\sigma_2:=\sigma_y=\Matrix{cc}{0&-{\rm i}\\ {\rm i}&0}\, ,\; 
\sigma_3:=\sigma_z=\Matrix{cc}{1&0\\0&-1}\, .
\eeq
Since here the local unitary group is $SU(2)^{\otimes n}$,
we only need fully local combs and hence can we restrict ourselves
to $SU(2)$ combs. We mention that any tensor product
$f(\{\sigma_\mu\}):=\sigma_{\mu_1}\otimes\dots\otimes\sigma_{\mu_n}$ 
with an odd number $N_y$ of $\sigma_y$ is an $n$-qubit comb. 
This can be seen immediately from
\nbeqa
(\!(f(\{\sigma_\mu\}))\!) &\equiv&
\bra{\psi} f(\{\sigma_\mu\}) \ket{\psi^*}=
\left (\bra{\psi^*} f(\{\sigma_\mu^*\}) \ket{\psi}\right )^*\\
&=& (-1)^{N_y} \bra{\psi} f(\{\sigma_\mu\})^\dagger \ket{\psi^*}=
	(-1)^{N_y}(\!(f(\{\sigma_\mu\}))\!)
\neeqa
In particular, 
\beq
       A^{(1)}_{1/2}:=\sigma_y {\mathcal C}
\eeq
is a comb acting on a single qubit.
We do not know, whether
combs acting on more than a single site might be needed to some extent.
As yet there is no evidence that they need to be included in order to classify 
multipartite qubit entanglement. 
In what follows, comb operators are to be understood
as acting on a single site only.

We will call 
$A^{(1)}: \mathfrak{h}\to\mathfrak{h}$ a comb of {\em order 1}.
In general we will call a (single-qubit) comb 
$A^{(n)}: \mathfrak{h}^{\otimes n}\to\mathfrak{h}^{\otimes n}$ to
be {\em of order n}.
It is worthwhile noticing that the $n$-fold tensor product
$\mathfrak{h}^{\otimes n}$ on which an $n$'th order comb acts symbolizes
$n$-fold copies of one single qubit state. In order to distinguish
this merely technical introduction of a tensor product of copies of states 
from the physically motivated tensor product of different qubits
we will denote the tensor product of copies with the symbol $\bullet$,
and hence write $A^{(n)}: \mathfrak{h}^{\bullet n}\to\mathfrak{h}^{\bullet n}$.
When we say {\em expectation value of } $A^{(n)}=L^{(n)}{\mathcal C}$ we mean
the following
\beq
\bra{\psi}^{\bullet n} A^{(n)} \ket{\psi}^{\bullet n}=:(\!(L^{(n)})\!)\; .
\eeq
Strictly speaking, this is a linear combination of products of expectation values:
if 
$$
A^{(n)}=\sum_{\vec{j}}a_{j_1,\dots,j_n}\sigma_{j_1}\bullet\dots\bullet\sigma_{j_n}{\mathcal C}
$$
then the {\em expectation value of }$A^{(n)}$ would be
$$ 
(\!(L^{(n)})\!)=\sum_{\vec{j}}a_{\vec{j}}\prod_{k=1}^n
\bra{\psi}\sigma_{j_k}\ket{\psi^*}=\sum_{\vec{j}}a_{\vec{j}}\prod_{k=1}^n(\!(\sigma_{j_k})\!)\; .
$$

There is a unique (up to rescaling) single site comb of order 2,
which is orthogonal to the trivial one $\sigma_y\bullet\sigma_y$ 
(with respect to the Hilbert-Schmidt scalar product):  
one can verify that for an arbitrary single qubit state
$$
0=(\!(\sigma_\mu \bullet \sigma^\mu)\!)= 
\langle \psi | \sigma_\mu{\mathcal C}| \psi \rangle
\langle \psi | \sigma^\mu{\mathcal C}| \psi \rangle := 
\sum_{\mu,\nu=0}^3 (\!( \sigma_\mu )\!)
		g^{\mu,\nu}(\!( \sigma_\nu )\!)\; ,
$$
with
$$ 
g^{\mu,\nu}=g_{\mu,\nu}=\Matrix{cccc}{
-1&0&0&0\\
0&1&0&0\\
0&0&0&0\\
0&0&0&1}\ ;\quad g^{0,0}=-1
$$
(notice the similarity with the Minkowski metric).
We will denote this second order comb by
\beq
A^{(2)}_{1/2}:=\sigma_\mu\bullet\sigma^\mu{\mathcal C}=
\sum_{\mu=0}^3 g^{\mu,\nu}\sigma_\mu\bullet\sigma_\nu\ {\mathcal C}\ \ .
\eeq

For both combs one can demonstrate that they are
$SL(2,\CC)$  invariant~\cite{OS05,DoOs08}, and hence they 
satisfy the basic requirements for the construction of filter operators.
Please note that any linear combination of combs is again a comb.
The two combs $A_{1/2}^{(1)}\bullet A_{1/2}^{(1)}$ 
and $A_{1/2}^{(2)}$ have the above-mentioned additional
important property of being mutually orthogonal with respect to 
the Hilbert-Schmidt scalar product.
Filter invariants for $n$ qubits are obtained as antilinear 
expectation values of filter operators;
the latter are constructed from combs so as to have vanishing 
expecation value for arbitrary product states.
We will use the word ``filter'' 
for both the filter operator and its filter invariant.
For $n$-qubit filters we will use the symbol ${\cal F}^{(n)}$. 

We start by writing down some filters for two qubits
\beqa\label{2-filters}
{\cal F}^{(2)}_1 &=&(\!(\sigma_y\otimes \sigma_y)\!)
 \\
{\cal F}^{(2)}_2 &=&\frac{1}{3}(\!([\sigma_\mu\otimes\sigma_\nu] \bullet
		  [\sigma^\mu\otimes \sigma^\nu])\!)=:\frac{1}{3}(\!(\sigma_\mu\sigma_\nu\bullet\sigma^\mu \sigma^\nu)\!)\label{2-filtersb}
\eeqa
Both forms are explicitly permutation invariant, and they are filters.
Indeed, if the state is a product, the combs lead to a vanishing expectation value.
We obtain the pure state concurrence from them in two different,
equivalent forms:

\beqa\label{2-measures}
C&=&\left |{\cal F}^{(2)}_1\right| \\
{\hspace*{-2mm}C^2}&=& 
                   \left|{\cal F}^{(2)}_2\right|
\eeqa

Now we go ahead to three qubits and write down a selection
of three-qubit filters
\beqa\label{3-filters}
{\cal F}^{(3)}_1 &=&(\!(\sigma_\mu\sigma_y\sigma_y\bullet
		  \sigma^\mu \sigma_y\sigma_y)\!)\\
{\cal F}^{(3)}_2 &=&\bigfrac{1}{3}
		(\!(\sigma_\mu\sigma_\nu\sigma_\lambda\bullet
		  \sigma^\mu \sigma^\nu\sigma^\lambda)\!)\\
{\cal F}^{(3)}_3 &=&
		(\!(\sigma_\mu\sigma_\nu\sigma_\lambda\bullet
		  \sigma^\mu \sigma_y\sigma_y\\
&& \qquad \bullet\sigma_y \sigma^\nu\sigma_y
	\bullet\sigma_y \sigma_y\sigma^\lambda)\!) \nonumber
\eeqa
The last two are evidently permutation invariant, but also
the first filter is invariant under permutations.
All coincide with the three-tangle~\cite{Coffman00} (or powers thereof) 
\beqa\label{3-measures}
\tau_3&=&\left |{\cal F}^{(3)}_1\right| 
      = \left|{\cal F}^{(3)}_2\right|\\
\tau_3^2 &=& 	\left |{\cal F}^{(3)}_3\right|
\eeqa
Interestingly, all two-qubit filters are homogeneous polynomials 
of the concurrence
and in the same way all three-qubit filters coincide with polynomials 
of the three-tangle. This is due to the fact that both concurrence 
and three-tangle generate the whole algebra of polynomial
$SL(2,\CC)^{\otimes 2}$ and $SL(2,\CC)^{\otimes 3}$ invariants,
respectively (see e.g.
Ref.~\cite{DoOs08} and references therein).

\subsection{Filters for 4 and more qubits}\label{unknownfilters}

In this section we will present explicitly a list of filters for systems of 
four and five qubits.  By means of the six-qubit example 
we sketch a straightforward procedure to construct filters for larger systems. 
In order to get compact formulas,
the tensor product symbol $\otimes$ will be omitted, as in eq.~\eqref{2-filtersb}.

For $4$ qubits, the whole filter ideal in the ring of polynomial $SL$ invariants
is generated by~\cite{DoOs08}
\beqa\label{4-filters}
{\cal F}^{(4)}_1 &=&
		(\!(\sigma_\mu\sigma_\nu\sigma_y\sigma_y\bullet
		  \sigma^\mu\sigma_y\sigma_\lambda\sigma_y
	\bullet\sigma_y\sigma^\nu\sigma^\lambda\sigma_y)\!) \\
{\cal F}^{(4)}_2 &=&\left\langle
		(\!(\sigma_\mu\sigma_\nu\sigma_y\sigma_y\bullet
		  \sigma^\mu\sigma_y\sigma_\lambda\sigma_y
	\bullet\sigma_y\sigma^\nu\sigma_y\sigma_\tau\right. \\
&& \qquad \left. \bullet
		\sigma_y\sigma_y\sigma^\lambda\sigma^\tau)\!)\right\rangle_s\nonumber\\
{\cal F}^{(4)}_3 &=&\bigfrac{1}{2}
		(\!(\sigma_\mu\sigma_\nu\sigma_y\sigma_y\bullet
		  \sigma^\mu\sigma^\nu\sigma_y\sigma_y
\bullet\sigma_\rho\sigma_y\sigma_\tau\sigma_y \bullet
		\sigma^\rho\sigma_y\sigma^\tau\sigma_y\nonumber\\
&&\qquad \bullet\sigma_y\sigma_\rho\sigma_\tau\sigma_y \bullet
		\sigma_y\sigma^\rho\sigma^\tau\sigma_y)\!)\nonumber\; .
\eeqa

For five qubits, we mention
\beqa\label{5-filters}
{\cal F}^{(5)}_{8;1} &=&\left\langle (\!(\sigma_{\mu_1}\sigma_{\mu_2}\sigma_{\mu_3}\sigma_2\sigma 
\bullet  \sigma^{\mu_1}\sigma^{\mu_2}\sigma_2\sigma_{\mu_4}\sigma_2 \bullet \right.\nonumber\\
&& \qquad \left.\sigma_{\mu_5}\sigma_2\sigma^{\mu_3}\sigma^{\mu_4}\sigma_2
  \bullet  \sigma^{\mu_5}\sigma_2\sigma_2\sigma_2\sigma_2)\!)\right\rangle_s\label{F1}\\
{\cal F}^{(5)}_{8;2} &=& \left\langle
(\!(\sigma_\mu \sigma_2  \sigma_2  \sigma_\nu  \sigma_\lambda \bullet    
\sigma^\mu \sigma_\rho \sigma_2 \sigma_2  \sigma^\lambda \bullet \right.    \nonumber\\
&&\qquad \left. \sigma_2 \sigma^\rho \sigma_\tau\sigma_2\sigma_\kappa \bullet     
 \sigma_2   \sigma_2   \sigma^\tau \sigma^\nu \sigma^\kappa)\!)\right\rangle_s, \label{F2}\\
{\cal F}^{(5)}_{8;3} &=& \left\langle 3
(\!(\sigma_\mu \sigma_\nu \sigma_\lambda \sigma_2 \sigma_2  \bullet         
\sigma_\tau \sigma^\nu \sigma^\lambda \sigma_2 \sigma_2 \bullet  \right.   \nonumber\\ 
&&\qquad \sigma^\tau\sigma_2\sigma_2\sigma_\rho \sigma_\kappa  \bullet      
\sigma^\mu  \sigma_2 \sigma_2 \sigma^\rho \sigma^\kappa)\!) \ +
\nonumber   \\
&& (\!(\sigma_\mu \sigma_\nu \sigma_\lambda \sigma_2 \sigma_2 \bullet    
\sigma^\mu \sigma^\nu \sigma^\lambda \sigma_2 \sigma_2 \bullet     \nonumber\\
&&\qquad \left.\sigma_\tau \sigma_2 \sigma_2 \sigma_\rho \sigma_\kappa \bullet    
\sigma^\tau \sigma_2 \sigma_2\sigma^\rho \sigma^\kappa)\!)\right\rangle_s. \label{T} \\
{\cal F}^{(5)}_0={\cal F}^{(5)}_{12;1} &=&(\!(\sigma_{\mu_1}\sigma_{\mu_2}\sigma_{\mu_3}\sigma_{\mu_4}\sigma_{\mu_5}
		   \bullet  \sigma^{\mu_1}\sigma_2\sigma_2\sigma_2\sigma_2 \bullet \\
&& \qquad\sigma_2\sigma^{\mu_2}\sigma_2\sigma_2\sigma_2
  \bullet  \sigma_2\sigma_2\sigma^{\mu_3}\sigma_2\sigma_2 \bullet \\
&& \qquad \sigma_2\sigma_2\sigma_2\sigma^{\mu_4}\sigma_2
  \bullet  \sigma_2\sigma_2\sigma_2\sigma_2\sigma^{\mu_5})\!) \nonumber \\
{\cal F}^{(5)}_{12;2} &=&\left\langle(\!(\sigma_{\mu_1}\sigma_{\mu_2}\sigma_{\mu_3}\sigma_2\sigma_2
	     \bullet \sigma^{\mu_1}\sigma^{\mu_2}\sigma_{\mu_4}\sigma_2\sigma_2 \bullet \label{5-12-2} \right.\\ 
&& \qquad  \sigma_{\mu_5}\sigma_2\sigma^{\mu_3}\sigma_2\sigma_{\mu_6} \bullet    
	     \sigma^{\mu_5}\sigma_2\sigma^{\mu_4}\sigma_2\sigma_{\mu_7}\bullet \nonumber\\
&& \qquad \left.\sigma_{\mu_8}\sigma_2\sigma_2\sigma_{\mu_9}\sigma^{\mu_6} \bullet 
		  \sigma^{\mu_8}\sigma_2\sigma_2\sigma^{\mu_9}\sigma^{\mu_7})\!)\right\rangle_{s;a}\nonumber\\
{\cal F}^{(5)}_{12;3} &=&\left\langle 
(\!(\sigma_{\mu_1} \sigma_{\mu_2}\sigma_{\mu_3}\sigma_2\sigma_2  \bullet 
\sigma^{\mu_1}  \sigma_2\sigma_{\mu_4}\sigma_{\mu_5}\sigma_2  \bullet 
\label{5-12-4}\right.\\
&& \qquad  \sigma_2 \sigma^{\mu_2}\sigma^{\mu_3}\sigma_2\sigma_{\mu_6} \bullet  
\sigma_{\mu_7}  \sigma_2\sigma^{\mu_4}\sigma_2 \sigma^{\mu_6} \bullet  
\nonumber\\
&& \qquad \left.\sigma^{\mu_7} \sigma_2\sigma_2 \sigma^{\mu_5}\sigma_{\mu_8} \bullet  
\sigma_2 \sigma_2 \sigma_2 \sigma_2 \sigma^{\mu_8})\!)\right\rangle_{s;a}\nonumber
\eeqa
\beqa
{\cal F}^{(5)}_{12;4} &=&\left\langle
(\!(\sigma_{\mu_1}\sigma_{\mu_2}\sigma_{\mu_3}\sigma_{2} \sigma_{2} \bullet 
\sigma^{\mu_1}\sigma^{\mu_2} \sigma_{\mu_4} \sigma_2\sigma_2  \bullet 
\label{5-12-G2}\right.\\
&& \qquad \sigma_{\mu_5}\sigma_{\mu_6}\sigma^{\mu_3}\sigma_2 \sigma_2  \bullet  
  \sigma^{\mu_5} \sigma_{\mu_7} \sigma_{2}\sigma_{\mu_8}\sigma_2 \bullet 
\nonumber\\
&& \qquad \sigma_{\mu_9}\sigma^{\mu_6} \sigma_2 \sigma^{\mu_8}\sigma_2  \bullet  
\sigma^{\mu_9} \sigma^{\mu_7}\sigma^{\mu_4}\sigma_2 \sigma_{2})\!) -\nonumber \\
&& 3
(\!(\sigma_{\mu_1}\sigma_{\mu_2}\sigma_{\mu_3}\sigma_{\mu_4} \sigma_{\mu_5} \bullet 
\sigma^{\mu_1}\sigma_2 \sigma_2 \sigma_2\sigma_2  \bullet 
\nonumber\\
&& \qquad \sigma_2\sigma^{\mu_2}\sigma_2\sigma_{\mu_6}\sigma_{\mu_7}  \bullet   
\sigma_{\mu_8}\sigma_{\mu_9}\sigma^{\mu_3}\sigma^{\mu_6}\sigma^{\mu_7} \bullet \nonumber\\
&& \qquad\left.\sigma^{\mu_8}\sigma^{\mu_9} \sigma_2 \sigma^{\mu_4}\sigma_2  \bullet 
\sigma_{2} \sigma_{2}\sigma_2\sigma_2 \sigma^{\mu_5})\!)\right\rangle_s\nonumber
\eeqa
where $\langle\dots\rangle_{s;a}$ means that the object between brackets
is symmetrized/antisymmetrized. Double indices indicate that both 
symmetrization and antisymmetrization lead to independent generators.
These filters can be found in Ref.~\cite{DoOs08}. 
Together with\footnote{
      Here, $P=\sum_j D_j$ and $D_j$ are those polynomial 
      $SL$-invariants of degree $4$ with a single $\sigma_\mu\bullet\sigma^\mu$ 
      contraction on qubit number $j$ as defined in Ref.~\cite{DoOs08}, e.g. 
      $D_1=(\!(\sigma_\mu\sigma_2\sigma_2\sigma_2\sigma_2\bullet 
               \sigma_\mu\sigma_2\sigma_2\sigma_2\sigma_2)\!)$; 
      they coincide with some invariants proposed in 
      Ref.~\cite{Wong00}.} 
$P^2-\sum_{j=1}^5 D_j^3$ 
and the square of an antisymmetric invariant $F$ 
that is constructed from an $\Omega$-process (see Ref.~\cite{Luque05}),
they generate the filter invariants for five qubits up to polynomial 
degree $12$.

The following examples for six-qubit filters provide the opportunity
to highlight a way to construct filters for higher qubit numbers 
\beqa\label{6-filters}
{\cal F}^{(6)}_1 &=&(\!(\sigma_{\mu_1}\sigma_{\mu_2}\sigma_y\sigma_y\sigma_y\sigma_y\bullet
		  \sigma^{\mu_1}\sigma_y\sigma_{\mu_3}\sigma_y\sigma_y\sigma_y\\
&& \qquad \bullet\sigma_{\mu_6}\sigma_y\sigma_y\sigma_{\mu_4}\sigma_y\sigma_y
 \bullet\sigma_y\sigma_y\sigma^{\mu_3}\sigma_y\sigma_{\mu_5}\sigma_y \nonumber \\
&& \qquad \bullet \sigma^{\mu_6}\sigma^{\mu_2}\sigma_y\sigma^{\mu_4}\sigma^{\mu_5}\sigma_y)\!)\nonumber
\\
{\cal F}^{(6)}_2 &=&(\!(\sigma_{\mu_1}\sigma_{\mu_2}\sigma_y\sigma_y\sigma_y\sigma_y\bullet
		  \sigma^{\mu_1}\sigma_y\sigma_{\mu_3}\sigma_y\sigma_y\sigma_y\\
&& \qquad \bullet\sigma_{\mu_6}\sigma^{\mu_2}\sigma^{\mu_3}\sigma_{\mu_4}\sigma_y\sigma_y
 \bullet\sigma_y\sigma_y\sigma_y\sigma^{\mu_4}\sigma_{\mu_5}\sigma_y \nonumber \\
&& \qquad \bullet \sigma^{\mu_6}\sigma_y\sigma_y\sigma_y\sigma^{\mu_5}\sigma_y)\!)\nonumber
\\
&\vdots& \nonumber \\
{\cal F}^{(6)}_i &=&(\!(\sigma_{\mu_\bullet}\sigma_{\mu_\bullet}\sigma_y\sigma_y\sigma_y\sigma_y\bullet
		  \sigma_{\mu_\bullet}\sigma_y\sigma_{\mu_\bullet}\sigma_y\sigma_y\sigma_y\\
&& \qquad \bullet\sigma_{\mu_\bullet}\sigma_\bullet \sigma_\bullet \sigma_{\mu_\bullet}\sigma_y\sigma_y
 \bullet\sigma_{\mu_\bullet}\sigma_\bullet \sigma_\bullet \sigma_\bullet 
\sigma_{\mu_\bullet}\sigma_y\bullet \sigma_\bullet \sigma_\bullet \sigma_\bullet 
\sigma_\bullet \sigma_\bullet \sigma_\bullet \bullet \dots \nonumber )\!)
\eeqa
where in the latter formula all the $\mu_\bullet$ are to be
contracted properly; in the $\sigma_\bullet$ the ``$\bullet$'' either 
have to be substituted by $\sigma_y$, or by indices which then have to be 
contracted properly. 
This suggests that for an $n$-qubit system the filter property requires
at least $\mathfrak{h}^{\bullet (n-1)}$, leading to polynomial degree
of at least $2(n-1)$ for the corresponding polynomial invariant.

We emphasize again that every filter is an $SL$
invariant because the local elements it is constructed from 
(i.e., the combs)
are $SL$ invariant. It is clear that linear combinations and in fact 
any function of invariants is an invariant
(but not necessarily an entanglement monotone, cf. Ref.~\cite{VerstraeteDM03}).
By noticing the consequences
of including global phases of the states we see that
only homogeneous functions of the same degree can be 
combined linearly.

\subsection{Maximally Entangled States}\label{maxent}

We will now define our notion of a
multipartite state with maximal genuine multipartite entanglement.
\begin{pdef}\label{def:maxent}
A pure $q$-qubit state $\ket{\psi_q}$ has maximal multipartite entanglement,
i.e.  q-tangle, {\em iff}
\begin{quotation} 
\item[(ia)]{The state is not a product, i.e.\
 the minimal rank of any reduced density matrix of $\ket{\psi_q}$ is $2$.}
\item[(ib)]{All reduced density matrices of $\ket{\psi_q}$ with rank
      $2$
      (this includes all $(q-1)$-site and single-site ones) 
      are maximally mixed within their range.}
\end{quotation}
{\em Further, there is a list of desirable features for 
maximally multipartite entangled states: }
\begin{quotation}
\item[(ii)]{all $p$-site reduced density matrices of $\ket{\psi_q}$,
      have zero $p$-tangle; $1<p<q$.}
\end{quotation}
{\em This is clearly an implicit requirement in that a check of it
would require the knowledge of convex-roof extensions of the relevant
multipartite entanglement measures. Furthermore, it is not even clear
which $q$-qubit entanglement families possess a representative for 
which all tangles for less than $q$ qubits vanish. The
$q$ qubit GHZ state is an example that satisfies condition (ii).}
\begin{quotation}
\item[(iii)]{there is a canonical form of any maximally $q$-tangled state, 
      for which properties (ia) and (ib) are unaffected by 
      relative phases in the amplitudes,  i.e.
      their quality of being maximally entangled is phase insensitive.}
\end{quotation}
\end{pdef}
All above requirements are invariant under 
local $SL$ transformations. We briefly discuss the implications of each
single requirement. 
Condition {\em (ia)} excludes product states, which are certainly not 
even globally $q$-tangled.
{\em (ib)} implies maximal gain of information 
when a bit of information is read out. This condition contains
the definition of {\em stochastic states} in Ref.~\cite{VerstraeteDM03},
where it is also proved that every entanglement monotone assumes its maximum
on the set of stochastic states. An even more stringent condition
has been imposed in Ref.~\cite{Gisin98}, where all reduced density matrices 
are required to being maximally mixed within their range. 
Requirements  {\em (ia)} and {\em (ib)} are therefore well-established. 
Constraint {\em (ii)} is intriguing by itself:
it excludes hybrids of various types of entanglement 
and thereby follows the idea of entanglement as a resource 
whose total amount has to be distributed 
among the possibly different types of entanglement; 
see e.g. Ref.~\cite{Coffman00,NoMonogamy}.
To our knowledge, it is not clear whether this condition can be regarded
as being fundamental, since up to date no extended monogamy relation
is found that would substantiate the idea of entanglement distribution
(see e.g.~\cite{Bai08,NoMonogamy}).  
We have no striking argument in favor of {\em (iii)}, except that maximally 
entangled states for two and three qubits have such a canonical form.  
We mention, however, that according to Ref.~\cite{Duer00} entangled states have 
a representation with a minimal number of product components; it appears that
in this representation the entanglement is not ``sensitive'' to changes in
the relative phases between the components (consider, e.g., the GHZ state).
It could be promising to analyze a possible connection to the concept
of {\em envariance} put forward in Ref.~\cite{Zurek-Env}.

In order to illustrate the above conditions and to
check the existence of such states, we give some examples:
The Bell states 
$(\ket{\sigma,\sigma'}\pm\ket{\bar{\sigma},\bar{\sigma'}})/\sqrt{2}$
are the canonical form of maximally $2$-tangled states. 
By tracing out one qubit one obtains $\frac{1}{2}\,\id$ as the reduced density matrix of the remaining qubit. 
The $2$-tangle is indeed robust against
multiplication of the components with arbitrary phases:
$(\ket{\sigma,\sigma'}+ e^{{\rm i}\phi}\ket{\bar{\sigma},\bar{\sigma'}})/\sqrt{2}$
is maximally entangled for arbitrary real $\phi$.  Condition  (ii) 
is meaningless here.
For two qubits, these are all maximally entangled states, and the class
of maximally entangled states is represented by
$\ket{GHZ_2}=\tfrac{1}{\sqrt{2}}(\ket{11}+\ket{00})$, 
which is like the GHZ state but for two qubits.
Also the generalized GHZ-state for $q$ qubits, 
$\tfrac{1}{\sqrt{2}}(\ket{1\dots 1}+\ket{0\dots 0})$,
satisfies all the above requirements.
It is straightforward to see that the GHZ state is
detected by every filter constructed in the way described in
the preceding sections.
Having a pure state of three qubits, there are two other classes of
entangled states:
the class represented by $\tfrac{1}{\sqrt{3}}(\ket{100}+\ket{010}+\ket{001})$,
which is equivalent to $\tfrac{1}{2}(\ket{000}+\ket{100}+\ket{010}+\ket{001})$
\cite{Duer00}, do not contain $3$-tangle at all. Indeed, they violate  
the requirements {\em (ib)} and {\em (ii)} in Def.~(\ref{def:maxent}).
An apparently different class of maximally $3$-tangled states instead 
can be read off directly from the coordinate expression for the 
three-tangle\cite{Coffman00}: its representative is
\beq\label{W-type}
\ket{X_3}=\tfrac{1}{2}(\ket{111}+\ket{100}+\ket{010}+\ket{001})
\eeq
and satisfies all the above conditions for a maximally $3$-tangled state. 
It is interesting to note that all its reduced two-site density matrices 
are an equal mixture of two orthogonal Bell states, which thus have 
zero concurrence. 
However, this state is unitarily
equivalent to a GHZ state by the transformation 
$H_2\otimes H_2\otimes H_2$, where  
$H_2$ is the Hadamard transformation
$\frac{1}{\sqrt{2}}\Matrix{cc}{1&1\\ 1&-1}$.
\\
Summarizing the above examples, we conclude that the set of states
that satisfy the conditions in Def.~(\ref{def:maxent}) is not empty for 
any number $q$ of qubits and we have one $2$-tangled and $3$-tangled 
representative (actually two, which are equivalent for three qubits, though).
In what follows we analyze the above conditions and prove that there 
are at least $q-1$ inequivalent $q$-tangled representatives.

\section{Polynomial SL(2,$\CC$) invariants and maximally entangled states}
\label{balancedness}

\subsection{The states measured by filters}
\label{eval}

In order to understand what the filters do, it is convenient to
consider subnormalized states
$$\ket{\psi}=:\sum_{j=0}^1\sum_{\{k_j\}} \psi_{j,\{k_j\}}\ket{j}\otimes\ket{\{k_j\}}
=:\sum_{j=0}^1\ket{j}\otimes\ket{\phi_j} $$
and then to study the action of the combs on this state.
Defining $\bar{0}:=1$ and $\bar{1}:=0$, we obtain
\beq\label{action:sigmay}
\bra{\psi}\sigma_2\otimes .\ket{\psi *} =\sum_j
        (-1)^{j+1} i \psi_{\bar{j},\{k_{\bar{j}}\}}
        \psi_{j,\{k_j\}}\bra{\{k_{\bar{j}}\}} . \ket{\{k_j\}}
\eeq

for the first order comb.
The second order comb is the sum of the following outcomes
\nbeqa\label{action:sigmamusigmamu}
-\bra{\psi}\sigma_0\otimes .\ket{\psi ^*}\bra{\psi}\sigma_0\otimes .\ket{\psi ^*} &=& 
      - \sum_{i,j}\bra{\phi_j} .\ket{\phi_j^*}\bra{\phi_i} .\ket{\phi_i^*}\\
\bra{\psi}\sigma_1\otimes .\ket{\psi ^*}\bra{\psi}\sigma_1\otimes .\ket{\psi ^*} &=& 
     \sum_{i,j} \bra{\phi_{\bar{j}}} .\ket{\phi_j^*}\bra{\phi_{\bar{i}}} .\ket{\phi_i^*}\\
\bra{\psi}\sigma_3\otimes .\ket{\psi ^*}\bra{\psi}\sigma_3\otimes .\ket{\psi ^*} &=& 
 \sum_{i,j} (-1)^{i+j} \bra{\phi_j} .\ket{\phi_j^*}\bra{\phi_i} .\ket{\phi_i^*}\; .
\neeqa
The remaining indices of the quantum state are kept fixed for the moment.
Summing up these three terms and performing the sum over $j$, we get
\beq
\sum_i 
\left[\bra{\phi_i} .\ket{\phi_{\bar{i}}^*}+\bra{\phi_{\bar{i}}} .\ket{\phi_i^*}\right]
  \bra{\phi_{\bar{i}}} .\ket{\phi_i^*}
-2 \bra{\phi_{\bar{i}}} .\ket{\phi_{\bar{i}}^*}\bra{\phi_i} .\ket{\phi_i^*}\; .
\eeq
It is seen from this result that in order to give a non-zero outcome,
every component $i$ must come with the flipped component $\bar{i}$.
This is what we will call a {\em balanced} qubit component.
Since the above consideration has to be extended to all qubits, 
we conclude that the filter has contributions only from the balanced 
part of a state.
We can say even more: the homogenous degree of the filter must fit with the 
length of the balanced  parts in the state, 
i.e. the number of product states in
the computational basis needed for this balanced part.
As a consequence, the way the filter is constructed already implies
valuable information about which type of state the filter 
can possibly detect.
This further underpins the relevance of polynomial
$SL(2,\CC)^{\otimes q}$ invariants as far as entanglement classification
and quantification of multipartite qubit states are concerned. 
In particular, we see that a state which can be locally transformed into a
normal form without balanced part has zero expectation value 
for all filters operators (the $\ket{W}$ state is a prominent example).

This analysis suggests in-depth investigation of states with balanced parts
in their pure-state decomposition into the computational (product) basis.
It is worth noticing, that it is not conclusive to look at
some given pure state and see whether it has a balanced part or not.
In fact, every pure state has a balanced part after a suitable choice of
local basis. The concept becomes useful only modulo 
local unitary transformations. Then, two qualitatively different classes of
states occur:
\begin{itemize}
\item states that are unitarily equivalent to a form without a balanced part,
\item states for which arbitrary local unitaries lead to a state 
  with a balanced part.
\end{itemize}
The latter case naturally splits up into two sub-classes. 
One is characterized as the reducibly balanced case in the 
sense that distinct smaller balanced parts always exist.
The complementary situation is the irreducibly balanced case.

It is clear that maximal entanglement as measured by some polynomial
SL-invariant is achieved when no unbalanced {\em residue} is present,
i.e. when the state is balanced as a whole. Indeed, we will show that
stochasticity of a state implies balancedness; 
against the background of the finding in Ref.~\cite{VerstraeteDM03} that
every entanglement monotone assumes its maximum on the set of stochastic 
states, this underpins a tight connection between balanced states 
and the notion of maximal (multipartite) entanglement.

\subsection{Irreducibly Balanced States}\label{irredbal}

For analyzing the first two conditions (ia) and (ib) in Def.~\ref{def:maxent} it
is convenient to express a pure state $\sum_i w_i \ket{i}$ as an array; 
the first row of this array contains the amplitudes $w_i$, $p_i:=|w_i|^2$.
The column below each amplitude is the binary sequence of the corresponding
product basis state. For example
$$
\tfrac{1}{2}(\ket{111}+\ket{100}+\ket{010}+\ket{001}) \longleftrightarrow
\Matrix{cccc}{
1/2 &1/2 &1/2 &1/2 \\
1& 1& 0& 0 \\
1& 0& 1& 0 \\
1& 0& 0& 1}\; .
$$
For the moment, we will not pay much attention to the weights
$p_i:=|w_i|^2$, they will become important later on (cf.\ Theorem III.3).
In the following, we define two types of matrices which are based 
on this array representation of a state. It will turn out that these
matrices are quite helpful in the discussion of the properties 
of balanced states. The proofs of some of the theorems will be
rather straightforward by using this representation.
\begin{defi}{alternating and binary matrix}
We call {\em binary matrix} $B_{\ket{\psi}}$ of the state $\ket{\psi}$ 
the matrix of binary sequences 
below the amplitude vector and equivalently we call
{\em alternating matrix} $A_{\ket{\psi}}$ of the state $\ket{\psi}$
the matrix obtained from its binary 
matrix, when all zeros are replaced by $-1$. It will be useful to allow for multiple repetition
of certain columns. This means of course that the alternating and binary matrix will not be unique.
The minimal form without repetitions is unique modulo permutations of the columns and qubits.\\
We define the {\em length} of a state as the minimal number of elements of
the standard product basis that occur in the state (without repetition of columns), i.e. the number 
of columns of the minimal form.
\end{defi}
In the above example we have
$$
B_{\ket{X_3}}=
\Matrix{cccc}{1& 1& 0& 0 \\
1& 0& 1& 0 \\
1& 0& 0& 1}\quad;
A_{\ket{X_3}}=
\Matrix{cccc}{1& 1& -1& -1 \\
1& -1& 1& -1 \\
1& -1& -1& 1}\; 
$$
and the length of this state is $4$. $A_{\ket{\psi}}$ and $B_{\ket{\psi}}$
are $q\times L$ matrices where $L$ is the number of basis states required for its
representation.
\begin{defi}{irreducibly balanced states}
\begin{enumerate}
\item We call a pure state $\ket{\psi}$ {\em (entirely) balanced} 
{\em iff} in each row of $B_{\ket{\psi}}$ there are as many ones as zeros 
(allowing for multiple occurrence of some of its columns),
or equivalently, {\em iff} the elements of each row of $A_{\ket{\psi}}$ 
sum up to zero (including multiplicities as for the binary matrix).
This can be expressed as
\beq\label{cond:balanced}
\exists\; n_1,\dots, n_L \in\NN (n_j>0) \ni 
\sum_{j=1}^L n_j (A_{\ket{\psi}})_{ij} = 0\; 
\forall\; i\in\{1,\dots,q\}
\eeq
where $A_{\ket{\psi}} \in \ZZ^{q\times L}$, i.e. qubit number $q$ and
length $L$.

We furthermore call a balanced state {\em irreducible} or 
{\em irreducibly balanced, iff} it cannot be split into  different smaller balanced parts
(i.e., iff there is no subset of less than L columns that is already balanced).\\
In contrast, a balanced state which can be split into different smaller balanced parts,
will be called {\em reducible}.
\item We call a pure state $\ket{\psi}$ {\em partly balanced} (i.e. it
has a balanced part) {\em if}~\eqref{cond:balanced} is satisfied for
$n_j\in\NN$ only if some $n_j=0$ (but not all of them).

A partly balanced state is called {\em reducible}/{\em irreducible}
{\em iff} its balanced part is reducible/irreducible.
\end{enumerate}
\end{defi}
As an example, we give the $B_{\ket{\psi}}$ matrix for a 
reducibly balanced 3-qubit state 
$$
B_{\mathrm{reducible}}=
\Matrix{cccc}{1& 0& 0& 1 \\
1& 0& 1& 0 \\
1& 0& 0& 1}\quad .
$$
\begin{defi}{completely unbalanced states}
We call a state 
{\em completely unbalanced}
if it is locally unitarily equivalent to 
a state without balanced part.
\end{defi}
Please note that the maximally $q$-tangled states for $q=2,3$ are
irreducibly balanced, and it can be straightforwardly verified that
they are the only ones for these cases. The W-states are completely unbalanced.
Further examples of completely unbalanced states are 
fully factorizing states. Therefore it is clear that 
complete unbalancedness can occur for both globally entangled
states and completely disentangled states; it therefore is an indicator only
as far as genuine multipartite entanglement (i.e., $q$-qubit entanglement
in $q$-qubit states) is concerned.

\begin{Theorem}
Product states are not irreducibly balanced.
\end{Theorem}
\begin{since}
First we observe that a product state is balanced {\em iff} its factors are.
Let the state be $\ket{\Phi}\otimes\ket{\Psi}$ which have $n$ and $m$ 
components, respectively, i.e. $\ket{\Phi}=\sum_{i=1}^n v_i \ket{\Phi_i}$
and $\ket{\Psi}=\sum_{i=1}^n w_i \ket{\Psi_i}$. Then 
$$
B_{\ket{\Phi}\otimes\ket{\Psi}}
\Matrix{ccccccc}{
B_{\ket{\Phi_1}} & \dots & B_{\ket{\Phi_1}} & \cdots & B_{\ket{\Phi_n}} & \dots 
& B_{\ket{\Phi_n}} \\
B_{\ket{\Psi_1}} & \dots & B_{\ket{\Psi_m}} & \cdots & B_{\ket{\Psi_1}} & \dots 
& B_{\ket{\Psi_m}} 
}
$$
with length $m n$ divided into $n$ blocks, 
$n>m$ without loss of generality.
Note that $m,n$ are even because $\ket{\Phi}$ and $\ket{\Psi}$
are assumed to be balanced. Consequently, the smallest common multiple of $m$ and $n$
is smaller than or equal to $\frac{m n}{2}$.
That is, there do exist $f,g\in\mathbb{N}$ relatively prime such that  
$f m = g n$ and $g\leq \frac{m}{2}$, $f\leq \frac{n}{2}$. 
Now we choose from each of the $n$ blocks ( corresponding to the state
$\ket{\Phi_i}\otimes\ket{\Psi}$, $i=1,\dots,n$ ) $g$ states such that
from the first $f$ blocks we choose the first state,
from blocks $f+1$ up to $2f$ we choose the 2nd state, ...,  
from blocks $(m-1)f + 1$ modulo $n$ up to $m\cdot f$ modulo $n$ we choose the $m$th state. 
This state is balanced and has length $m\cdot f\leq m n/2$. 
This proves that any product state - if balanced - is reducible.
\end{since}
It is important to emphasize that every state can be made balanced 
by local unitary transformations, 
raising in general the number of components in the state.
\begin{Theorem}
Every balanced $q$-qubit state
with length larger than $q+1$ is reducible.
\end{Theorem}
\begin{since}
Balancedness of the state implies the existence of $n_1,\dots,n_L$ such that
$\sum_{j=1}^L n_j (A_{\ket{\psi}})_{ij} = 0$.
Irreducibility implies that no subset ${\cal K}$ of ${\cal L}:=\{1,\dots,L\}$
exists with ${\cal K}\cap{\cal L}\neq {\cal L}$ such that 
$\sum_{j\in{\cal K}} m_j (A_{\ket{\psi}})_{ij} = 0$ for some positive integers
$m_j$.  Without loss of generality $A_{\ket{\psi}}$ has rank $q$.
In order to fix the idea of the proof, we insert a vertical cut in the
matrix $A_{\ket{\psi}}$ such that both parts contain at least $(q+1)/2$ colums.
This means to introduce two disjunct non-empty sets ${\cal K}$ and 
${\cal K}':={\cal L}\setminus  {\cal K}$ with 
$|{\cal K}|,|{\cal K}'|\geq \frac{q+1}{2}$.
We define $\vec{\alpha}^{\cal K}=:(\alpha^{\cal K}_1,\dots,\alpha^{\cal K}_q)$ 
such that $\alpha^{\cal K}_i=\sum_{j\in{\cal K}} n_j (A_{\ket{\psi}})_{ij}$.
Irreducibility implies $\vec{\alpha}^{\cal K}\neq \vec{0}$.
We now split nonempty sets $\kappa\subset{\cal K}$ and $\kappa'\subset{\cal K}'$
off the subsets ${\cal K}$ and ${\cal K}'$
and define $\tilde{\cal K}:=({\cal K}\setminus \kappa) \cup \kappa'$.
Then, including arbitrary non-negative integers $m_j$, $j\in\kappa'$, and
defining $m_j=n_j$ for $j\in {\cal K}$, we find
$$
\sum_{j\in\tilde{\cal K}} m_j  (A_{\ket{\psi}})_{ij}
= \alpha^{\cal K}_i - \sum_{j\in\kappa} m_j (A_{\ket{\psi}})_{ij}
+\sum_{j\in\kappa'} m_j (A_{\ket{\psi}})_{ij}\; .
$$
Irreducibility then implies that for all such subsets ${\cal K}$ and 
$\kappa$ no integer numbers $\tilde{m}_j$ ($\tilde{m}_j$ can be also negative or zero)
do exist such that $\vec{\tilde{m}}\in\ZZ^{|\kappa|+|\kappa'|}$ satisfies the condition
$\sum_{j\in\kappa\cup\kappa'}\tilde{m}_j (A_{\ket{\psi}})_{ij}=\alpha^{\cal K}_i$
for all $i\in\{1,\dots,q\}$. 
Without loss of generality we can assume that 
$(A_{\ket{\psi}})_{i\in\{1,\dots,q\};j\in\kappa\cup\kappa'}$ has rank $q$
(a suitable choice of ${\cal K}$ and $\kappa$ guarantees that). This
implies that the condition can be satisfied for every integer vector $\vec{\alpha}^{\cal K}$
even in $\ZZ^q$, hence contradicting our assumption of irreducibility.
\end{since}
A comment is in order here. It must be stressed that the integers
entering the balancedness condition must be positive. 
Therefore, linear dependence of the column vectors does not 
imply balancedness. In fact, not every $q$ qubit state with more than $q+1$ product state 
components is balanced. The reason is that the set of positive integers
is not a field. Our proof however nicely highlights that the balancedness 
condition itself bridges this gap and provides a mapping onto a set of linear 
equations over the field $\ZZ$. Thus, for balanced states 
the argument of linear independence can  indeed be used.
The state being irreducibly balanced thus implies that the rank
of its corresponding alternating $(q\times L)$-matrix ($q$ rows
and $L$ columns)  is equal to $L-1$. Since its maximal rank is $\min \{q,L\}$,
this implies $L\leq q+1$. A canonical form of such an irreducibly 
balanced state thus becomes
\beq\label{irreducibly-balanced}
\Matrix{cccccccc}{
0 & 0 & \dots & 0 & 1 & 1 & \dots & 1 \\ 
0 & 0 & \dots & 1 & 0 & 1 & \dots & 1 \\ 
\dots & \dots & \dots & \dots & \dots  & \dots & \dots & \dots \\ 
0 & 1 & \dots & 0 & 0 & 1 & \dots & 1 \\ 
1 & 0 & \dots & 0 & 0 & 1 & \dots & 1 
}\; .
\eeq
>From this canonical form, further such states (except the GHZ state) 
can be generated by duplicating rows, NOT-operations on certain
bits, and permutation of rows, i.e. of bits.
We mention that in Ref.~\cite{NoMonogamy}, the procedure of duplicating
rows has been termed {\em telescoping}; it was used to generate certain 
multipartite entangled states that obey a monogamy relation.
 
\begin{Theorem}\label{stochastic->balanced}
Stochastic states (i.e. states satisfying (ia) and (ib)) are balanced.
In other words: stochasticity implies balancedness.
\end{Theorem}
\begin{since}
For the proof, let us consider an arbitrary $q$-qubit state $\ket{\psi}$ 
that satisfies the conditions (ia), (ib) 
- and maybe (iii) - in Def.~(\ref{def:maxent}) 
and trace out everything but the first qubit. 
We can write the array of the state $\ket{\psi}$ as follows
\beq\label{state}
\Matrix{cccccc}{
w_1 &\dots & w_n & w_{n+1}&\dots&w_L \\
1& \dots& 1& 0& \dots& 0 \\
\Phi_1& \dots& \Phi_n& \Phi'_1& \dots& \Phi'_{L-n}
}\; .
\eeq
Now assume that some of the states $\ket{\Phi_i}$ coincide with some of the
$\ket{\Phi'_i}$, and call $\ket{\Psi}$ their superposition with corresponding weights
of the right hand side;  the corresponding superposition of the $\Phi_i'$
can be written as $\ket{\tilde{\Psi}}=\alpha\ket{\Psi} + 
\beta\ket{\Psi_\perp}$, $|\alpha|^2+|\beta|^2=1$;
$\braket{\tilde{\Psi}}{\tilde{\Psi}}=:\tilde{x}$. 
Note that these states are not normalized to one.
Then, the whole state can be written as
$(\ket{0} +\alpha \ket{1})\otimes \ket{\Psi}
            +\beta \ket{1}\otimes \ket{\Psi_\perp}
 + \ket{0}\otimes \ket{\Psi'}  + \ket{1}\otimes \ket{\Psi''}$
with pairwise orthogonal states $\ket{\Psi}$, $\ket{\Psi_\perp}$, 
$\ket{\Psi'}$, and $\ket{\Psi''}$ and squared norms $x$, $x_\perp$, $y$, 
and $z$ respectively.
The one-site reduced density matrix then is
$\Matrix{cc}{ \tilde{x} +z & x \alpha \\ 
              x \alpha^* & x + y }$. 
Since the only $2\times 2$ density matrix
with two eigenvalues $\frac{1}{2}$ is $\frac{1}{2} \id $, the condition (ib) of definition~\ref{def:maxent}
 implies: ($x=0$, $y=\frac{1}{2}$) or ($\alpha=0$, $y=\frac{1}{2}-x$).
The latter relation means that $\ket{\Psi}\perp\ket{\tilde{\Psi}}$,
and gives a phase-relation for the amplitudes $w_i$, contradicting
the requirement (iii).
The former condition means that no two
$\ket{\Phi_i}$, $\ket{\Phi'_i}$ are equal and therefore orthogonal.
In both cases (irrespective the phase insensitivity) we find
\beq\label{1st}
\rho^{(1)}=\Matrix{cc}{
\sum\limits_{i\in I^j_1} p_i & 0 \\
0 & \sum\limits_{i\in I^j_0} p_i}\stackrel{!}{=}
\frac{1}{2}\id\quad \Leftrightarrow\quad 
\sum_{i\in I^j_1} p_i - \sum_{i\in I^j_0} p_i = 0
\eeq
where $I^j_1$, $I^j_0$ are the set of column numbers, where the $j$-th qubit 
has value $1$ and $0$ respectively.
The above applies to each single qubit.
Eq.(\ref{1st}) can be written in a more compact form in terms of the 
corresponding alternating matrix and the vector 
$\vec{p}:=(p_1,\dots,p_L)$ of weights:
\beq\label{eq:balanced}
A_{\ket{\psi}} \vec{p} =\vec{0}
\eeq
Note that balancedness means
\beq
A_{\ket{\psi}} \vec{1} =\vec{0}
\eeq
where $\vec{1}:=(1,\dots,1)$.
Equation~\eqref{eq:balanced} has a unique solution with all weights equal {\em iff}
the state is irreducible\footnote{Allowing that one
state can occur more than once, corresponding to a higher weight}.
Otherwise the state is reducible and all states in each irreducible block
$B_b$ have the same weight $p_b$. 
This corresponds to a superposition of irreducibly balanced states.
Phase insensitivity however turns out to be incompatible with more
than one block except when all the states $\ket{\Phi_i}$, $\ket{\Phi'_j}$
were perpendicular to each other. 
This means that the superposed irreducibly balanced states must be orthogonal
to each other. 
Tracing out only one qubit (including possible telescope copies of it)
gives exactly the same condition~(\ref{1st}).
\end{since}

It is worth noticing at this point that with local operations on $q$ qubits, 
the maximum number of free phases is $q+1$ (including a global phase), 
which coincides with the maximum length of an irreducibly 
balanced block.
Therefore the only remaining reducible states which could be 
maximally entangled by virtue of the demanded phase insensitivity 
are to be superpositions of irreducible ones with 
total length not larger than that of the irreducible state of maximal length.

We want to remark that 
we can - without loss of generality - shrink all states such 
that no telescope bits occur; the shrinking does not affect the reducibility.

The above observations eventually lead to the following set of 
maximally entangled $q$-qubit states of maximal length
\beq\label{Max-Entangled}
\ket{X_q}=\sqrt{q-2}\ket{1\dots 1}+\sum_i \ket{i}
\eeq
where $\ket{i}$ denotes the state with all bits zero except the $i$-th,
which is one. The maximally entangled state of minimal length is 
always the GHZ state.
States of intermediate length are obtained from those with maximal length for
$p$ qubits ($p<q$) by means of 
telescoping\footnote{see discussion after (\ref{irreducibly-balanced})}.

It is worth mentioning that irreducibility does not trivially imply
the form~\eqref{irreducibly-balanced}. An example for such a 
state of five qubits
is
\beq\label{eq:exception}
B_{\chi}=\Matrix{cccccc}{1&1&1&0&0&0\\1&1&0&1&0&0\\
1&0&1&0&1&0\\1&0&1&0&0&1\\1&0&0&0&1&1}
\eeq

\begin{Theorem}\label{irred-bal:equiv:stochastic}
Every irreducibly balanced state is equivalent under 
local filtering operations $SL^{\otimes q}$ (LFO) to
a stochastic state.
\end{Theorem}
\begin{since}
The proof goes by construction.
Let $a_j$, $j=1,\dots,L$ be the amplitudes of the product state
written in the $i$-th column of $B_{\ket{\psi}}$, and
let us consider the LFO's
\beq\label{LFOcomplex}
{\cal T}^{(i)}_{\rm LFO}=\Matrix{cc}{t_i &0\\ 0 & t_i^{-1}}
\eeq
with $t_i=:t^{z_i}$ for some real positive $t$ and complex $z_i$; $i=1,\dots,q$.
Without loss of generality let the multiplicities $n_j=1$
for all (at most $q+1$) $j$ 
(differring multiplicities could be absorbed in the $p_i$).
We must then show that after suitable LFO's all amplitudes
are equal.\\
After this transformation, the amplitude of the product states
(i.e. of the column vectors) would become 
\beq
a_j t^{\sum_{i=1}^q z_i (A_{\ket{\psi}})_{ij}} =: a_j t_j\; .
\eeq
Balancedness implies $\prod_{j=1}^L t_j = 1$.
Without loss of generality let $(B_{\ket{\psi}})_{i,1}=0$ 
for all $i=1,\dots,q$.
Dividing by $t_1$ the amplitudes become
$a_j t^{\sum_{i=1}^q 2z_i(B_{\ket{\psi}})_{ij}}$. Stochasticity requires that all amplitudes
have to be equal (up to a phase) and leads to the set of linear equations
$$
\sum_{i=1}^q 2z_i(B_{\ket{\psi}})_{ij} = \log_{t} \frac{a_1}{a_j}\; ;\
j=2,\dots,L\;.
$$
Since $L\leq q+1$ and $B_{\ket{\psi}}$ has rank $L-1$, a solution
vector $(z_1,\dots,z_q)$ exists for arbitray $a_j\neq 0$.
The resulting pure state is stochastic. 
\end{since}
The fact that every irreducibly balanced state is SL-equivalent to a stochastic state,
in combination with the negation of theorem~\ref{stochastic->balanced},
leads to the following
\begin{Corollary}\label{not-unbalanced}
An irreducibly balanced state is locally unitarily inequivalent to every
state without balanced part. In other words, irreducibly balanced states
are not completely unbalanced. 
\end{Corollary}
In the light of the fact that the minimal number of orthogonal product states
in which a pure quantum state can be represented is invariant
under $SL(2,\CC)$ transformations~\cite{Duer00}, 
the following property of irreducibly
balanced states is important.
\begin{Theorem}\label{minimality}
For $q>3$ qubits, irreducibly balanced states descending from
Eq.~\eqref{Max-Entangled} are {\em minimal} in the 
sense that an irreducibly balanced state of length $L$ can not be 
represented as a superposition of less than $L$ states of a
computational basis (i.e., elements of  a completely factorized basis).
\end{Theorem}
\begin{since}
The irreducibly balanced state $\ket{X_q}$ of $q$ qubits (cf.\
eq.\ (\ref{Max-Entangled})) has length $q+1$ and its
$(q-1)$ qubit reduced density matrix is spanned by a generalized 
$(q-1)$ qubit GHZ and W state; for $q>3$ it has no product state in its range. 
The minimal lengths of the $(q-1)$ qubit GHZ and W are $2$ and $(q-1)$ 
respectively,
and hence they are different for $q\neq 3$ (this implies that they are 
SLOCC-inequivalent~\cite{Duer00}). 
It can be shown that the possibility to express $\ket{X_q}$ as a 
superposition of less than $q+1$ computational basis 
states implies that there must
be a product state in the range of the $(q-1)$ qubit reduced density
matrix, which leads to a contradiction.
This inductively proofs the minimality of all irreducibly balanced
states as defined before.  
\end{since}

It is an important step now to realize the following
\begin{Theorem}
All irreducibly balanced states belong to the SLOCC non-zero class,
i.e. they are robust against infinitely many 
LFO's $SL^{\otimes q}$ and therefore 
possess a finite normal form~\cite{VerstraeteDM03}.
As a consequence, they are maximally entangled states according to 
definition~\ref{def:maxent} (also in the sense of Ref.~\cite{VerstraeteDM03}).
\end{Theorem} 
\begin{since}
Those transformations that go beyond $SU(2)$ are essentially the 
LFO's of the form
\beq\label{LFO}
{\cal T}^{(i)}_{\rm LFO}=\Matrix{cc}{t_i &0\\ 0 & t_i^{-1}}
\eeq
when expressed in a suitable local basis for the $i$-th qubit.
Now assume the existence of LFO's that scale the state down
to zero after infinitely many applications - we will call this
the {\em zero-class assumption}.
Defining a set of real numbers $p_i\in\RR$, $i=1,\dots,q$ such that
$t_i=:t^{p_i}$ with $t>1$ (without loss of generality), 
the action of this single LFO rescales the weight of the 
$j$-th column of the alternating matrix $A_{\ket{\psi}}$ by the factor
\beq
t^{s_j}=t^{\sum_{i=1}^q p_i A_{ij}}\; ;\quad \mbox{such that }
s_j<0\mbox{ for all }j\in\{1,\dots,L\}
\eeq
where $L$ is the length of the state of $q$ qubits and the negativity of
all the $s_j$ expresses the zero-class assumption.~\footnote{It is important to realize in this context that every finite 
succession of LFO's of the type~\eqref{LFO} can be replaced by a single LFO
with suitably chosen set of real numbers $\tilde{p}_i$.}
This is equivalent to
\beq\label{scalable}
0>w_j= \sum_{i=1}^q p_i A_{ij}
\eeq
for all $j\in\{1,\dots,L\}$.
Now we make use of the balancedness of the state, meaning that
$A_{i,L}=-\sum_{j=1}^{L-1} A_{i,j}$ for all $i\in\{1,\dots,q\}$
and that \eqref{scalable} must apply to all $j\in\{1,\dots,L-1\}$
by virtue of our zero-class assumption.
Consequently
\nbeqa
w_L &=& \sum_{i=1}^q p_i A_{iL}\\
&=& -\sum_{j=1}^{L-1}\sum_{i=1}^q p_i A_{i,j}\\
&=& -\sum_{j=1}^{L-1} w_j >0\; .
\neeqa
Thus, at least one positive scaling exponent must exist.
This contradicts our initial assumption.\\ 
Now it is crucial that for irreducibly balanced states 
no basis exists in which the state has no balanced part 
(theorem~\ref{not-unbalanced}). This completes the proof.
\end{since}
The same applies to $q$ qubit states which are
superpositions of orthogonal irreducibly balanced states with length 
smaller than $q+2$.

As a consequence, there must exist independent entanglement monotones 
which attribute to each of these states a non-zero value, and which 
can distinguish between them.
Equivalently, all completely unbalanced states belong to the 
SLOCC zero-class.
An example is the class of $W$ states for arbitrary number of qubits, but also 
products of states where at least one of the factors is part of
the corresponding $SLOCC$ zero-class. Therefore, every $SL(2,\CC)^{\otimes q}$
invariant entanglement monotone when applied to such states gives zero. 

We now briefly discuss the requirements in definition~\ref{def:maxent} 
for irreducibly balanced states with particular emphasis on condition (ii) of vanishing sub-tangles.
The GHZ state satisfies all conditions; the subtangles are
all trivially zero, because the reduced density matrices are mixtures
of product states. Therefore, they are maximally $q$-tangled.
The states of maximal length behaves analogously,
except that tracing out the first qubit,
a mixture of a generalized GHZ state and a W state is obtained.
For three qubits, the resulting W state is a GHZ (or Bell-) state and the mixture has zero 
$2$-tangle.
Also for four qubits one can construct a decomposition of the density matrix
whose elements all have zero $3$-tangle. For growing number of qubits,
the GHZ weight monotonically decreases to zero.
However, it can be shown that it
contains a subtangle that is detected by certain factorized filters.

It is straightforward to show that the GHZ state is detected by all
simple filters  (that is those $SL(2,\CC)$ invariants that are directly created
from invariant combs, but not e.g. linear combinations of such invariants).

\section{Toward a generalization beyond qubits}\label{gen}

\subsection{Compound entanglement or block entanglement}\label{compounds}

The 
invariant-comb approach also provides suggestions 
how to possibly extend  
such an ansatz towards entanglement measures for blocks of spins
of variable size.
To this end we exploit the fact that each operator with an odd number of 
$\sigma_y$ is a comb. Furthermore, if two $q$-qubit filters
are identical for pure $q$-qubit states, but not identical as operators, 
their difference is a comb. Examples are
$\sigma_\mu \sigma_i\bullet  \sigma^\mu \sigma_i - \sigma_\mu \sigma_j\bullet  \sigma^\mu \sigma_j $, for $i\neq j$ fixed, and
$\sigma_\mu \sigma_\nu\bullet  \sigma^\mu \sigma^\nu - 
          3 \sigma_y \sigma_y\bullet \sigma_y \sigma_y$ for two qubits
and $\sigma_\mu \sigma_\nu \sigma_\tau \bullet  \sigma^\mu \sigma^\nu \sigma^\tau
- 3 \sigma_\mu \sigma_y \sigma_y \bullet  \sigma^\mu \sigma_y \sigma_y$
for three qubits.
However, 
this constitutes just a starting point as this
will typically lead to a set of combs on which the local unitary
group acts in a non-trivial way; in order to guarantee that a constructed filter
is an entanglement monotone, we need an {\em invariant} comb.
Clearly, abandoning the requirement for the monotone property
would open up a vast variety of possible ``measures'' or ``indicators'' 
for entanglement. This, however, is not what we have in mind.
 
In order to be invariant, it is necessary that the combs are regular and 
all their eigenvalues must have equal modulus. 
This is a clear criterion for designing an approach we have in mind. 
The approach pursued
e.g. in~\cite{Heydari05} has some overlap with 
concurrence vector approaches (see Refs.~\cite{Wootters01,Akhtarshenas05}),
which for bipartite systems coincides with the 
universal state inversion (see Ref.~\cite{Wootters01} and Ref.~\cite{Heydari05}).
The local antilinear operators used there are not regular
and therefore can not be invariant under local $SL$ operations in 
higher local dimensions.

This opens up a rich and promising field for future investigation.
Some insight into the intrications and consequences
involved with this requirement is given in the next section
on general spin $S$. 
It is worth noticing that the concept of balancedness introduced above
is tailor-made for qubit systems; it is not appropriate for higher local
dimension. The notion of maximal entanglement would need to be modified
correspondingly, once such invariant combs have been found.

\subsection{Spin $S>\bigfrac{1}{2}$}\label{spinS}

The operator $S_y=-{\rm i} (S^+-S^-){\cal C}$ is a comb for arbitrary spin $S$. 
The crucial difference to the spin-1/2 case is that
there are more first-degree combs for $S>1/2$ due to the fact that there 
are non-trivial powers of spin operators up to order $2 S>1$ -- 
since $(S^+)^{2S+1}=0$.
It turns out that for $S=1$, there is a three-parameter 
variety of first-degree combs
\beq\label{spin1Toeter}
A_{1}[a,b,c]:= \left(a S_y + ( b S_x S_y + c S_y S_z + {\rm h.c.})\right){\cal C}
\eeq
and a six-parameter variety for spin $\bigfrac{3}{2}$
\beq\label{spin3/2Toeter}
A_{3/2}[a,b,c]:= S_y \left( a \id +  b S_x + c S_z 
                       + d  S_x S_z + e S_x^2 + f S_z^2 + {\rm h.c.}\right ){\cal C}
\eeq
As in the qubit case, every product of spin operators 
containing an odd number of $S_y$ (plus its hermitean conjugate)
is a comb.
A generalization to general spin $S$ is therefore straight forward:
We have $S(2S+1)$ independent off-diagonal (pure imaginary) entries,
which is the dimension of the variety.
The corresponding operators are those appearing in $(S_x+S_z)^m$; $m=0,\dots,2 S-1$.

Unfortunately, for integer spin, i.e. for $2S+1$ odd, these combs
are not regular. This follows from the {\em hairy ball theorem},
stating that in order to have a continuous map from the surface
of a $d$-dimensional sphere onto itself, $d$ has to be even.
In our case the surface corresponds to the real part of
the normalized Hilbertspace (due to the antilinearity, 
every comb on the real Hilbert space is a comb on all the Hilbert space). 
Therefore, for integer spin, one has to look out for a comb of higher order.
Unfortuantely, also for half-integer spin no first order  
$SL(2S+1,\CC)$ invariant combs do exist.

In order to make a first step towards higher spins 
in the spirit of the invariant-comb approach, 
let us first consider a simplified scenario, 
where only local rotations are accessible in the laboratory.
Then, the group of local operations is the complex extension of 
the $2S+1$ dimensional representation of $SU(2)$, hence still $SL(2,\CC)$. 
We want to stress that this situation differs considerably
from that of an arbitrary $2S+1$ level system, where the most
general local operations are out of the complexified $SU(2S+1)$, 
which is the $SL(2S+1,\CC)$.
For half-integer spin $S$, the $SL(2,\CC)$-invariant comb is 
obtained as
$$
A_S={\rm i}\; {\rm antidiag}\{(-1)^m\,;\ m=1,\dots,2S+1\}\,{\cal C}\; ;
\quad 2S \quad {\rm odd}
$$
with associated linear operator
$$
L_S={\rm i}\; {\rm antidiag}\{(-1)^m\,;\ m=1,\dots,2S+1\}\; .
$$
Here, ${\rm antidiag}\{\lambda_1,\dots,\lambda_n\}$ indicates the 
$n\times n$ matrix with $\lambda_1, \dots,\lambda_n$ on the anti-diagonal,
e.g. $\sigma_y={\rm antidiag}\{-{\rm i},{\rm i}\}$.
With these combs, we can immediately construct an analogue for the concurrence
for arbitrary half-integer spin
\beq\label{Concurrence-SpinS}
C_S[\Psi]=\left |\bra{\Psi} L_S\otimes L_S \ket{\Psi^*}\right |
\eeq
for which the convex-roof extension procedure from Ref.~\cite{Uhlmann}
can be applied, and hence the $SL(2,\CC)$-concurrence for general 
half-integer spin $S$ is
\beqa\label{ConcS}
C_S &=& \max\{0,2 \lambda_1 - \trace R\} \\
&& \quad R=\sqrt{\sqrt{\rho} L_S\otimes L_S \rho^* L_S\otimes L_S \sqrt{\rho}}\nonumber
\eeqa
It must be stressed that this concurrence is a measure of entanglement
under restricted local operations, namely to local rotations of the
cartesian axis. The notion of SLOCC is modified correspondingly.
Each restricted entanglement class will be subdivided
into classes with respect to the full group of local 
transformations $SL(2S+1,\CC)$.
We therefore are confident that an analysis of the
$SL(2,\CC)$ invariant concurrence (\ref{ConcS}) will nevertheless
give interesting insight into the entanglement classes
for higher local dimensions.

It could be interesting to compare these combs with 
further existing proposals as 
the {\em universal state inversion}\cite{Rungta}, which however are 
constructed for general $d$-state systems.
We leave this investigation for future studies.  

\section{Conclusions}

In the recent literature, an efficient procedure for the construction 
of local $SL(2,\CC)^{\otimes q}$ invariant operators for $q$ qubit 
wavefunctions has emerged out of  the simple requirement to
create entanglement indicators that should vanish for
{\em all} product states (a minimal requirement for a quantity 
to detect only global entanglement)~\cite{OS04,OS05}.
We call this procedure the {\em invariant-comb approach},
because the local building blocks already are $SL(2,\CC)$ invariant.
It is interesting that some definitely globally entangled states as the W state
are not detected by any of these polynomial invariants. This motivates the
concept of {\em genuine multipartite entanglement} in order to distinguish
globally entangled quantum states detected by some non-zero polynomial
{\em SL} invariant from others. 
The fact that those invariants automatically lead to 
entanglement monotones has motivated our detailed analysis of the
properties of many-qubit states that are detected by the
entanglement measures created from invariant combs.
We have chosen an approach from two different points of view, 
with significantly overlapping results: 
On the one hand, we find that a necessary requirement for a pure 
quantum state of many qubits to have finite genuine multipartite 
entanglement is that the state has a balanced part.
This balancedness constitutes a continuation of the curious geometric
picture of the three-tangle as highlighted in Ref.~\cite{Coffman00} to a higher 
number of qubits.
On the other hand, also basic necessary requirements for 
maximal pure-state entanglement, namely that the state has to
be stochastic~\cite{Gisin98,VerstraeteDM03}, are demonstrated here
to readily imply balancedness.
This curious coincidence justifies a systematic analysis of balanced states.
We have extracted the locally $SU(2)$ invariant
``nucleus'' of balancedness, which is the {\em irreducible balancedness}.
It is shown that irreducibly balanced states are locally {\em SL} invariant to
stochastic states, a prerequisite for being maximally entangled.
Irreducible balancedness is also shown to exclude the existence of 
a completely unbalanced form (as e.g. the W state has).
This result is essential in that it demonstrates that irreducibly balancedness
is a well-defined and valuable concept.
Furthermore we could prove that irreducibly balanced states belong
to the non-zero SLOCC class of states; hence, they have a non-trivial
normal form after local filtering operations~\cite{VerstraeteDM03}.

A canonical form for a family of irreducibly balanced states has been found,
and this family has the minimal number of components in 
a fully factorized basis.
This minimal ``length'' is a {\em non}-polynomial 
{\em SL} invariant~\cite{Duer00,Lamata07} which according to
our analysis has 
a tight connection with an entanglement classification using
{\em polynomial} {\em SL} invariant entanglement measures.
This connection consists in that the homogeneous degree of the polynomial 
invariant has to fit with the length of the balanced part of the minimal form.
From the latter we can read off (up to a normalization factor)
the value of the polynomial SL invariant.

Precise sufficient conditions have been singled out for reducibly balanced states
in order to be maximally entangled. Such states clearly exist,
possibly even without irreducibly balanced form.
However, irreducibly balanced states provide
a generating ``basis'' (not claiming completeness) for the costruction of such states,
in the sense that reducibly balanced states are superpositions of irreducibly balanced ones.

It is worth giving reference to a collective entanglement measure proposed in~\cite{Wallach}.
For qubits, it is equivalent to the averaged one-tangle 
$\tau_1=\frac{4}{n}\sum_j \det \rho_j$ (see e.g. Ref.~\cite{Barnum03}), where 
$\rho_j$ is the reduced density matrix of qubit number $j$.
These measures are only sensitive to the requirement (ia) of Def.~\ref{def:maxent}. 
They assume their global maxima for all those maximally entangled states presented here
(satisfying the condition (ib)).
This also includes arbitrary tensor products of such maximally entangled states.
So, this measure is an indicator of stochasticity of a pure state, but cannot
discriminate any of the SLOCC entanglement classes present in that state.
This shortcoming might be overcome to some extent by looking at maxima of 
suitable functions of e.g. von Neumann entropies of certain reduced density matrices.
Such an analysis has been pursued among others in Ref.~\cite{Love06} 
and has singled out the four qubit ``X state'' in eq.~(\ref{Max-Entangled}): 
an irreducibly
balanced state in the canonical form presented here (and before in Ref.~\cite{OS04,OS05}).

An additional advantage of the invariant-comb approach is that it suggests possible generalization to
general subsystems. We have discussed to some extent generic complications encountered with
such an extension. A specific analysis for bipartite entanglement of general half-integer 
spins is added. 
With a  restriction of the local operations
to just local rotations in the laboratory, an analogue to the concurrence is presented 
explicitly and its exact convex-roof extension has been constructed using a result of 
Ref.~\cite{Uhlmann}. Its comparison with other existing proposals remains to be
further investigated.

\acknowledgments
The authors would like to thank L. Amico, D.\ \DJo, C. Eltschka, 
R. Fazio, P. Horodecki, A. Uhlmann, and O.\ Viehmann
for helpful discussions. This work was supported by the EU 
RTN {\em Nanoscale Dynamics, Coherence, and Computation}, 
grant HPRN-CT-2000-00144 and by the German Research Foundation within
Sonderforschungsbereich 631 and the Heisenberg Programme.

%\bibliography{biblio}

\end{document}